# Phenomenology from 100 large lattices*

Tanmoy Bhattacharya and Rajan Gupta [a]

[a]T-8 Group, MS B285, Los Alamos National Laboratory, Los Alamos, New Mexico 87545 U. S. A.

We present a status report on simulations being done on $32^3 \times 64$ lattices at $\beta = 6.0$ using quenched Wilson fermions. Phenomenologically relevant results for the spectrum, decay constants, the kaon B-parameter $B_K$, $B_7$, $B_8$, semi-leptonic and $B \to K^*\gamma$ form factors are given based on a statistical sample of 100 configurations.

## 1. LATTICE PARAMETERS

The $32^3 \times 64$ gauge lattices were generated at $\beta = 6.0$ using the combination of 5 over-relaxed (OR) and one Metropolis or Pseudo-heatbath sweep. We have stored lattices every 2000 OR sweeps. Quark propagators, using the simple Wilson action, have been calculated with periodic boundary conditions in all 4 directions for two kinds of extended sources – Wuppertal and Wall – at $\kappa = 0.135$ ($C$), 0.153 ($S$), 0.155 ($U_1$), 0.1558 ($U_2$), and 0.1563 ($U_3$). These quarks correspond to pseudoscalar mesons of mass 2816, 977, 687, 541 and 428 MeV respectively where we have used $1/a = 2.314$ GeV for the lattice scale. The three light quarks allow us to extrapolate the data to the physical isospin symmetric light quark mass $\overline{m}$, while the $C$ and $S$ $\kappa$ values are selected to be close to the physical charm and strange quark masses.

We analyze three types of hadron correlators distinguished by the type of source/sink used to generate quark propagators. These are (i) wall source and point sink (WL), (ii) Wuppertal source and point sink (SL), and Wuppertal source and sink (SS). The effective mass $m_{eff}(t)$ converges to the asymptotic value from below for WL correlators and from above for SL and SS correlators in all hadron channels. At the final $2\sigma$ level the convergence is extremely slow and there exist correlated fluctuations lasting $3 - 10$ time-slices. As shown in Fig. 1, the pion signal persists far enough to confirm that WL and SL give the same mass. In all cases where we can make consistency checks, we find that the mean of the $WL$ and $SL$ mass is a very good estimate for the asymptotic value. These estimates are presented in Table 1.

Figure 1. Comparison of the convergence of $m_{eff}(t)$ for WL and SL pion ($U_1U_1$) correlators.

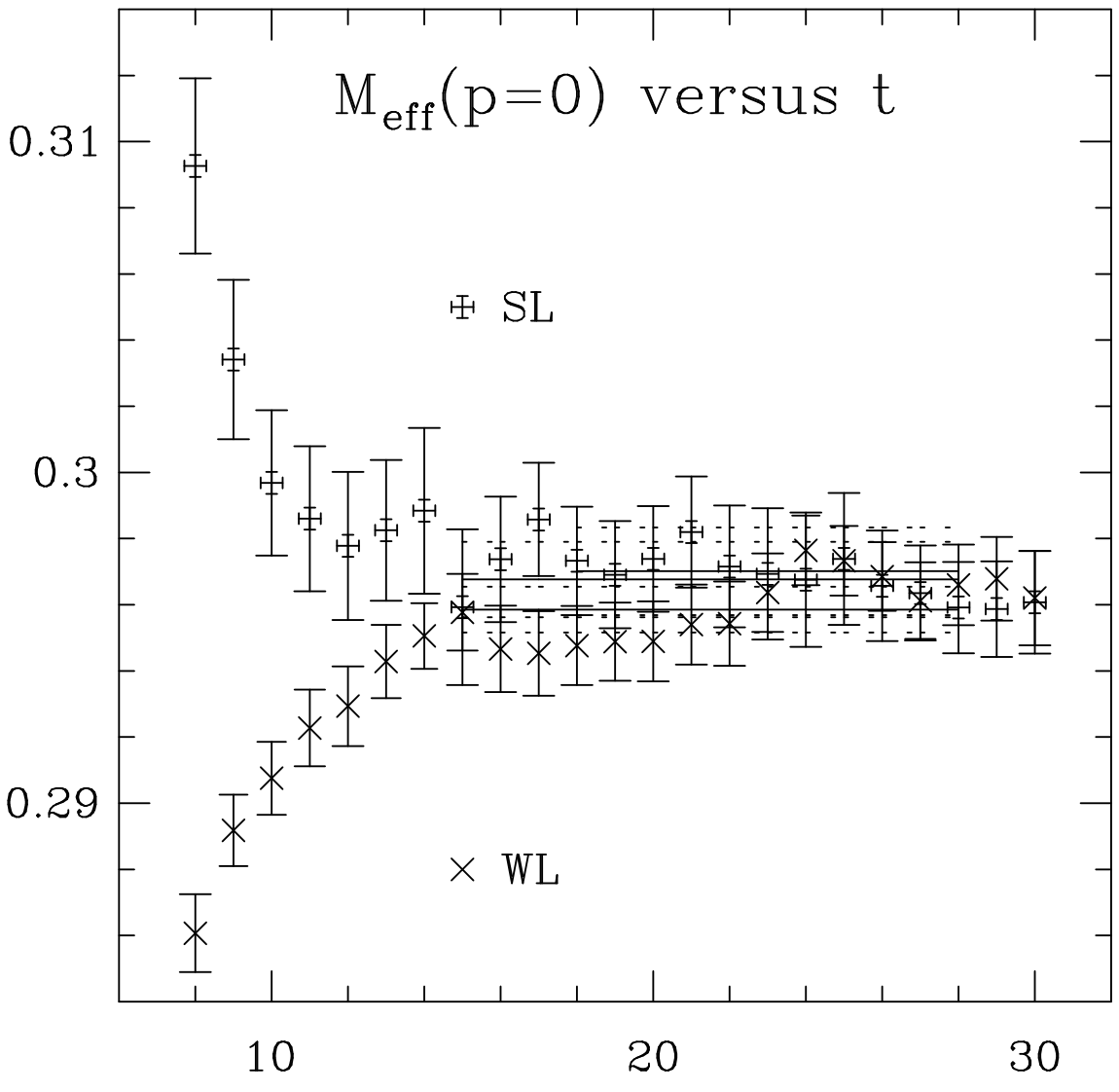

Errors are calculated using a single elimination Jackknife procedure with all necessary fits and extrapolations done independently for each JK sample. Possible artifacts due to the quenched approximation are not included in the analysis but are discussed in the review talk by Gupta [2].

To extrapolate the data to the physical quark masses, we verify that $m_K^2$, $m_\pi^2$, $m_\rho$, and $m_\phi$ are linear in the quark mass $m_q$ defined as

*Based on talks presented by Tanmoy Bhattacharya and Rajan Gupta. These calculations have been done on the CM5 at LANL as part of the DOE HPCC Grand Challenge program, and at NCSA under a Metacenter allocation.

$0.5(1/\kappa - 1/\kappa_c)$ or determined non-perturbatively as discussed in Ref. [1]. The results are essentially identical since the two estimates of $m_q$ are linearly related to high accuracy. An extrapolation of the average of WL, SL, and SS pion masses to zero gives $\kappa_c$, and of the ratio $m_\pi^2/m_\rho^2$ to the physical value 0.03182 gives $\kappa_l$ corresponding to $\overline{m}$:

$$\kappa_c = 0.15714(1), \qquad \kappa_l = 0.15705(1). \qquad (1)$$

The lattice scale, as determined from $m_\rho$, is

$$a^{-1}(m_\rho) = 2.314(74) \text{ GeV}. \qquad (2)$$

We determine $\kappa_s$ in three ways. We extrapolate $m_K^2/m_\pi^2$, $m_{K^*}/m_\rho$, and $m_\phi/m_\rho$ to $\overline{m}$ and then interpolate in the strange quark to match their physical value. Using the procedure described in [2], $m_s$ evaluated at 2 GeV in the $\overline{MS}$ scheme is

$$\kappa_s(M_K) = 0.1550(1), \; m_s(M_K) = 129(04) \text{ MeV}$$
$$\kappa_s(M_{K^*}) = 0.1547(3), \; m_s(M_{K^*}) = 151(15) \text{ MeV}$$
$$\kappa_s(M_\phi) = 0.1546(3), \; m_s(M_\phi) = 157(13) \text{ MeV}$$

All results for matrix elements are presented using local operators. We use the Lepage-Mackenzie improvement scheme in estimating the perturbative renormalization factors [3]. The "boosted" coupling is defined as

$$\frac{1}{g^2_{\overline{MS}}(q^* = \pi/a)} = \frac{\langle plaq \rangle}{g^2_{latt}} + 0.025. \qquad (3)$$

## 2. SPECTRUM

The lattice dispersion relation for hadrons deviates from the continuum form $E^2 = p^2 + m^2$ due to discretization errors, and is not known *a priori* for bound states due to nonperturbartive effects. This can have important consequences for the calculation of matrix elements involving heavy quarks. We compare four simple ansätze for the dispersion relation in Fig. 2 for $CU_1$. The data favor the nearest neighbor symmetric difference relativistic dispersion relation $\sinh^2(E/2) - \sin^2(p/2) = \sinh^2(m/2)$, which implies that the "kinetic" mass $m_2^{-1} \equiv (\partial^2 E/\partial p^2|_{p=0})^{-1} = \sinh m$. A comparison of $m_1$ (given by the rate of exponential fall off) and this $m_2$ is given in Table 1 for the $\pi$ and $\rho$ mesons, and the data show a significant difference (due to $O(ma)$ effects) for the charm mesons.

Figure 2. Test of four different dispersion relations for $CU_1$ pion correlator

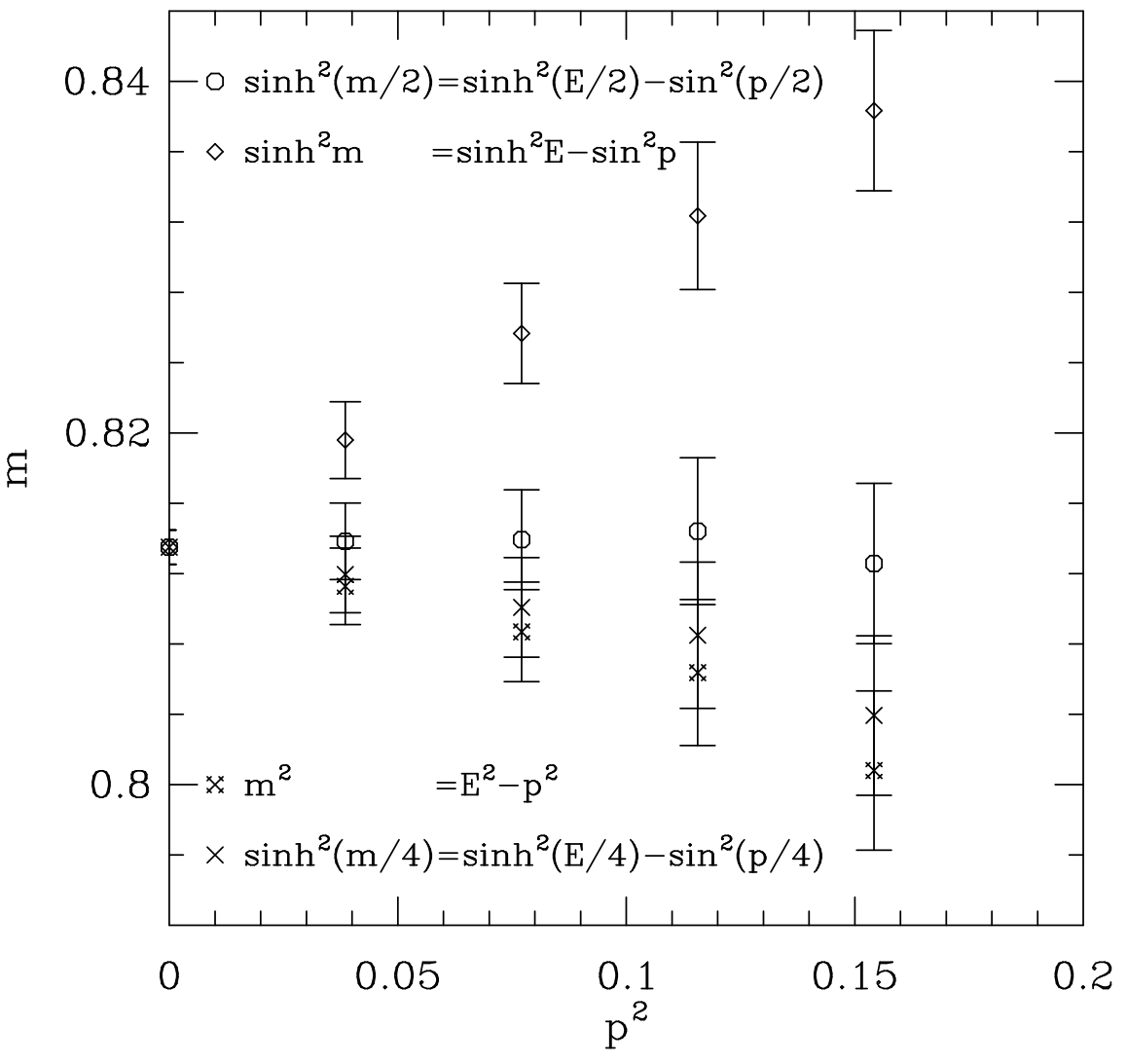

### 2.1. Baryons

The signal in wall and Wuppertal baryon correlators becomes noisy at roughly the point where the $m_{eff}$ begin to converge as exemplified in Fig. 3. We therefore use the average of $SL$ and $WL$ correlators as our best estimate. These numbers are given in Table 2 for the case of degenerate quarks. Extrapolating the data for the 3 light quarks to $\overline{m}$ gives

$$m_N a = 0.482(13); \qquad m_\Delta a = 0.590(30). \qquad (4)$$

Comparing these results with GF11 data [5] we find that our value for $M_N/m_\rho$ lies roughly $1.5\sigma$ higher while that for $M_\Delta/m_\rho$ is consistent.

The non-degenerate data is in excellent agreement with the Gell-Mann-Okubo formula $2(M_N + M_\Xi) = 3M_\Lambda + M_\Sigma$ if either $S$ or $U_1$ are designated as the strange quark and any of the $U_i$ as the light quark. Similarly, the $SU(6)$ breaking mass difference $m_\Delta - m_N$ is determined to be 289(42) MeV, in good agreement with the experimental value of 293 MeV. On the other hand, the two independent splittings in the nucleon multiplet turn out to be significantly lower than their experimental values (using $a^{-1}(m_\rho)$ and $m_s(m_\phi)$):

$$m_{\Sigma_N} - m_N = 185(17) \text{ MeV cf. expt. } 254 \text{ MeV}$$
$$m_{\Xi_N} - m_N = 300(27) \text{ MeV cf. expt. } 379 \text{ MeV},$$

Table 1. Meson masses in lattice units

| | $m_\pi$ | | $m_\rho$ | |
|---|---|---|---|---|
| | $m_1$ | $m_2$ | $m_1$ | $m_2$ |
| $CC$ | 1.217(1) | 1.541(2) | 1.229(1) | 1.564(2) |
| $CS$ | 0.854(1) | 0.962(2) | 0.880(1) | 0.999(2) |
| $CU_1$ | 0.814(1) | 0.906(2) | 0.841(1) | 0.945(3) |
| $CU_2$ | 0.799(1) | 0.886(3) | 0.827(2) | 0.924(3) |
| $CU_3$ | 0.790(2) | 0.873(4) | 0.817(2) | 0.911(5) |
| $SS$ | 0.422(1) | 0.435(1) | 0.506(1) | 0.530(3) |
| $SU_1$ | 0.364(1) | 0.372(1) | 0.465(2) | 0.482(4) |
| $SU_2$ | 0.339(1) | 0.347(1) | 0.449(2) | 0.464(4) |
| $SU_3$ | 0.324(1) | 0.331(2) | 0.439(3) | 0.453(5) |
| $U_1U_1$ | 0.297(1) | 0.302(1) | 0.422(3) | 0.435(5) |
| $U_1U_2$ | 0.267(1) | 0.271(2) | 0.405(4) | 0.416(7) |
| $U_1U_3$ | 0.248(1) | 0.251(2) | 0.394(5) | 0.404(8) |
| $U_2U_2$ | 0.234(1) | 0.237(2) | 0.387(5) | 0.396(9) |
| $U_2U_3$ | 0.211(1) | 0.214(2) | 0.376(7) | 0.382(12) |
| $U_3U_3$ | 0.185(1) | 0.187(2) | 0.363(9) | 0.365(15) |

Table 2. Baryon masses in lattice units

| | $SSS$ | $U_1U_1U_1$ | $U_2U_2U_2$ | $U_3U_3U_3$ |
|---|---|---|---|---|
| $N$ | 0.789(03) | 0.641(04) | 0.579(06) | 0.540(12) |
| $\Delta$ | 0.835(04) | 0.706(08) | 0.660(13) | 0.631(27) |

and $\approx 20\%$ lower still if $m_s(m_K)$ is used instead. Similarly, $m_\Omega - m_\Delta = 322(31)$ MeV (cf. expt. 440 MeV).

## 3. DECAY CONSTANTS

There are many different ways of combining $SL$ and $SS$ correlators to extract $f_{PS}$ and $f_V$, some of which are described in Ref. [1]. We find that all methods give essentially identical results, and the data for $f_{PS}$ is independent of the momentum, a necessary condition for the reliable extraction of semi-leptonic form factors. We quote the mean value in Table 3 for two common renormalization schemes, naive ($\sqrt{2\kappa}$) and Lepage-Mackenzie tadpole improved (imp). The large dependence of $f_{PS}$ and $f_V$ on the scheme for heavy quarks due to $O(ma)$ corrections underscores the need for determining the renormalization constant accurately. The results below are with the "imp" scheme.

We find $f_\pi = 141(6)$ MeV, roughly $2\sigma$ *larger* than the physical value 130.7 MeV, in contrast to the result from GF11 collaboration [5]. Using $f_\pi$ to set the scale we find $a^{-1}(f_\pi) = 2.162(62)$ GeV. Similarly, we find that $f_K = 162(5)$ MeV ($f_K = 166(5)$ MeV) with $m_s$ set by $M_K$ ($M_\phi$) as compared to the physical value 160 MeV. Some of the systematic errors cancel in the ratio $f_K/f_\pi$, which we determine to be 1.15(1) (1.18(2)).

Figure 3. $WL$ and $SL$ effective mass versus $t$ for nucleon and $\Delta$ using $U_1U_1U_1$ quarks.

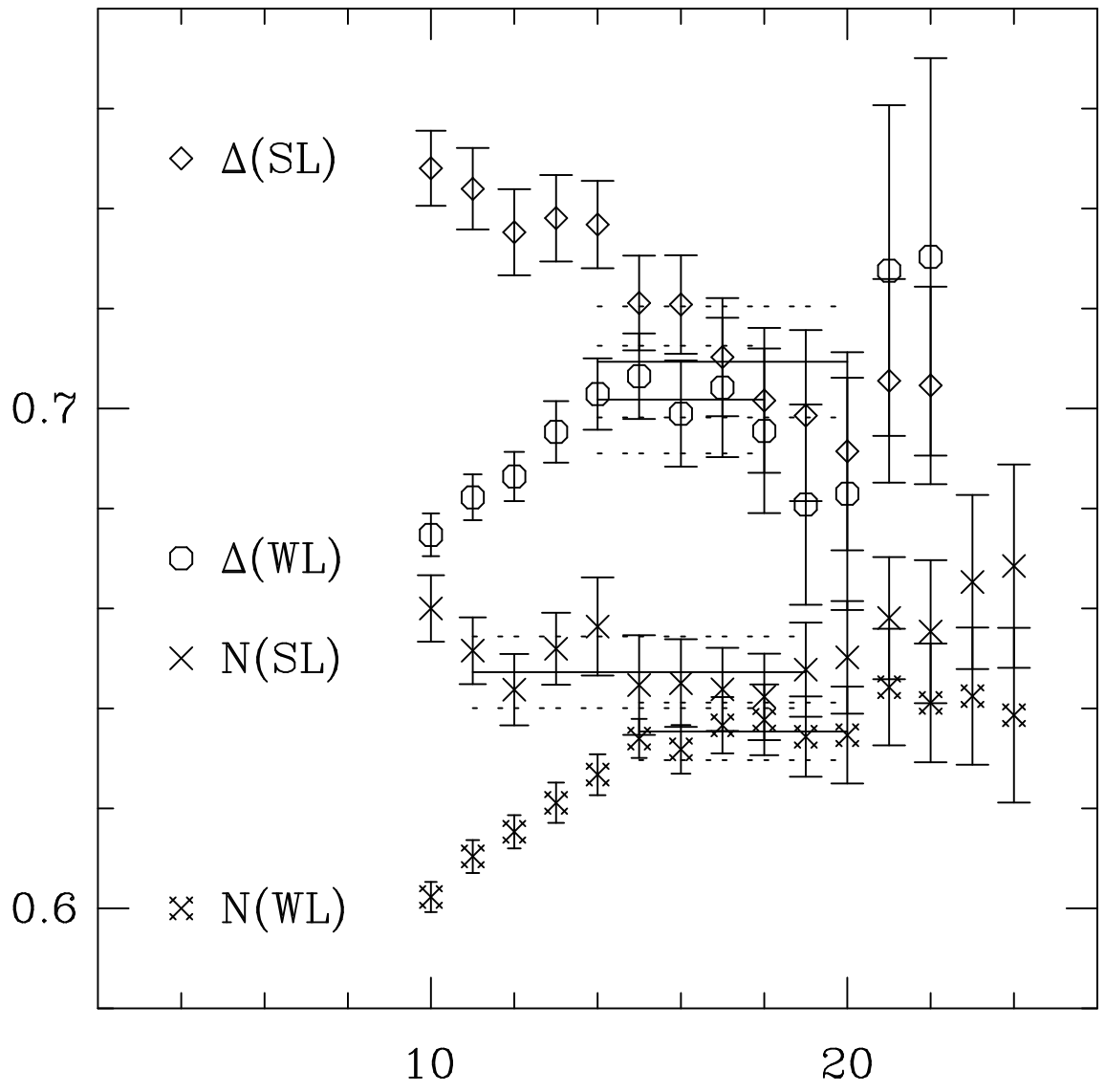

The $D$ meson with our heavy $C$ quark has $m_1 = 1789(56)$ and $m_2 = 1972(62)$ MeV using $a^{-1}(m_\rho)$, and about 7% lower using $a^{-1}(f_\pi)$. The physical value $m_D = 1869$ MeV lies within the range of these systematic uncertainties. Therefore, we take $C$ to be the physical charm quark, and present results for $f_D$ and $f_{D_s}$ in Table 4.

The data in Fig. 4 show that $f_V^{-1}$ is qualitatively different for degenerate versus non-degenerate quark masses. We extrapolate the 4 lightest degenerate combinations to $\overline{m}$ to get $f_\rho^{-1} = 0.357(8)$, somewhat smaller than the experimental value 0.398 and consistent with previous determinations [5]. For mesons involving strange and charm quarks we get :

$$f_{K^*}^{-1} = 0.326(12), \qquad f_\phi^{-1} = 0.317(7),$$
$$f_{D_s^*}^{-1} = 0.194(4)\ , \qquad f_{D^*}^{-1} = 0.167(6).$$

Using $m_s(m_K)$ reduces these by less than $1\sigma$.

Table 3. $f_\pi$ and $1/f_V$ in lattice units

| | $f_\pi^{naive}$ | $f_\pi^{imp.}$ | $1/f_V^{naive}$ | $1/f_V^{imp.}$ |
|---|---|---|---|---|
| $CC$ | 0.125(2) | 0.202(3) | 0.104(2) | 0.173(3) |
| $CS$ | 0.100(2) | 0.132(2) | 0.129(2) | 0.176(3) |
| $CU_1$ | 0.091(2) | 0.118(2) | 0.125(3) | 0.166(3) |
| $CU_2$ | 0.088(2) | 0.112(2) | 0.122(3) | 0.161(4) |
| $CU_3$ | 0.086(2) | 0.109(3) | 0.120(4) | 0.157(5) |
| $SS$ | 0.089(1) | 0.096(1) | 0.254(4) | 0.282(5) |
| $SU_1$ | 0.083(1) | 0.087(1) | 0.268(5) | 0.289(5) |
| $SU_2$ | 0.080(1) | 0.084(1) | 0.271(6) | 0.290(6) |
| $SU_3$ | 0.079(1) | 0.081(2) | 0.272(6) | 0.289(7) |
| $U_1U_1$ | 0.077(1) | 0.079(1) | 0.291(5) | 0.307(5) |
| $U_1U_2$ | 0.074(1) | 0.076(1) | 0.298(6) | 0.311(6) |
| $U_1U_3$ | 0.073(1) | 0.073(1) | 0.302(6) | 0.310(7) |
| $U_2U_2$ | 0.071(1) | 0.072(1) | 0.308(7) | 0.314(7) |
| $U_2U_3$ | 0.070(1) | 0.070(1) | 0.311(8) | 0.315(8) |
| $U_3U_3$ | 0.068(1) | 0.067(1) | 0.312(9) | 0.315(9) |

Table 4. $f_D$ and $f_{D_s}$ in MeVwith scale from $m_\rho$.

| | $m_s(M_K)$ | | $m_s(M_\phi)$ | |
|---|---|---|---|---|
| | $m_1$ | $m_2$ | $m_1$ | $m_2$ |
| $f_D$ | 240(10) | 229(10) | | |
| $f_D/f_\pi$ | 1.70(4) | 1.62(4) | | |
| $f_{D_s}$ | 272(08) | 258(08) | 279(06) | 264(06) |
| $f_{D_s}/f_D$ | 1.13(2) | 1.13(2) | 1.16(3) | 1.15(3) |

## 4. $B_K$, $B_7$, $B_8$ with Wilson fermions

The methodology used to calculate the $B$-parameters is presented in [6] [4], so we only summarize the new results here. The calculation is done with the two operators $\bar{\psi}\gamma_5\gamma_4\psi$ and $\bar{\psi}\gamma_5\psi$ to create and annihilate the kaons. The convergence of the matrix elements is from opposite directions for these two cases and thus provides a check on the results. We find that the two sets of data are essentially identical except for $U_2U_3$ and $U_3U_3$ combinations for the $LL$ operator. The final values are taken to be the mean of the two cases and given in Table 5[2]. All data are obtained using 1-loop improved operators as defined in Ref. [6] and evaluated with $g^2_{boost.} = 1.616$ and $\mu a = \pi$.

$B_K$: The chiral behavior of the matrix element

[2]The numbers presented in Table 3 in LAT93 are wrong. The actual data is consistent with the present analysis.

Figure 4. $f_V^{-1}$ versus $m_q$.

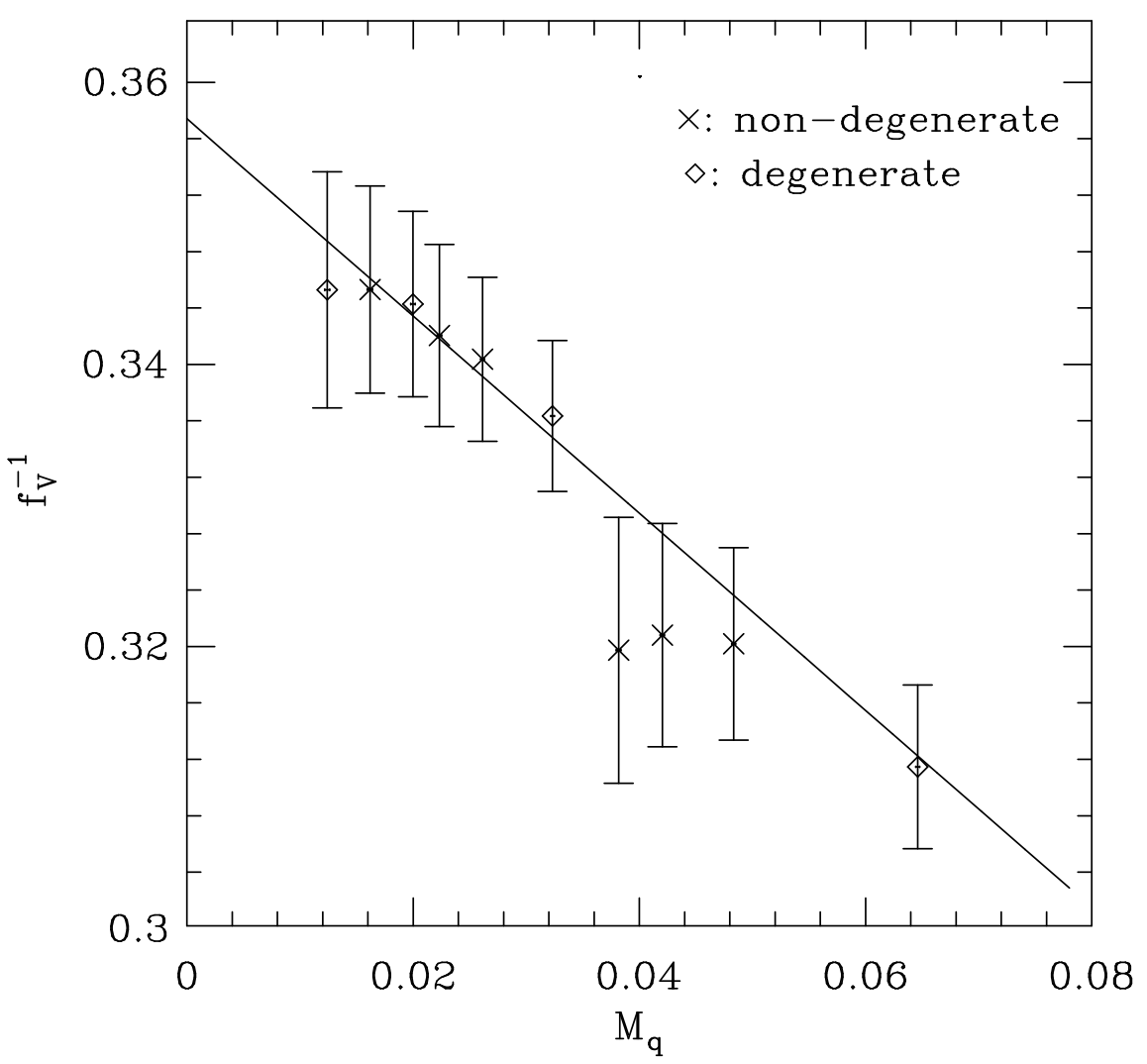

of the $LL$ operator (ignoring chiral logs) is

$$\left\langle \overline{K^0} \right| \mathcal{O}_{LL} \left| K^0 \right\rangle = \alpha + \beta m_K^2 + \gamma p_i p_f + \delta_1 m_K^4 + \delta_2 m_K^2 p_i p_f + \delta_3 (p_i p_f)^2 + \ldots \quad (5)$$

A fit to the data for the lightest 10 mass combinations is shown in Fig. 5. (Very similar values for the six parameters are obtained from fits to 4 or 6 lightest mass combinations.) We find that even though the statistical quality of the data is very good (see Table 5), the three $\delta_i$ are not well determined; only $\delta_2$ is significantly different from zero. Terms proportional to $\alpha, \beta$ and $\delta_1$ are pure lattice artifacts due to the bad chiral behavior induced by the mixing of the $\Delta S = 2$ 4-fermion operator with wrong chirality operators. The coefficients $\gamma, \delta_2, \delta_3$ contain artifacts in addition to the desired physical pieces and we do not have a way of resolving these two contributions. We simply assume that the 1-loop improved operator does a sufficiently good job of removing these residual artifacts. Using 1-loop matching between lattice and continuum gives $B_K$ at 7.26 $GeV$ to be

$$B_K = \gamma + (\delta_2 + \delta_3) m_K^2 = 0.67(11) \quad (6)$$

a value consistent with the much more precise staggered fermion data [7]. A second way of extracting $B_K$ is to use Eq. 5 for pairs of points

Figure 5. Fit to the $B_K$ data.

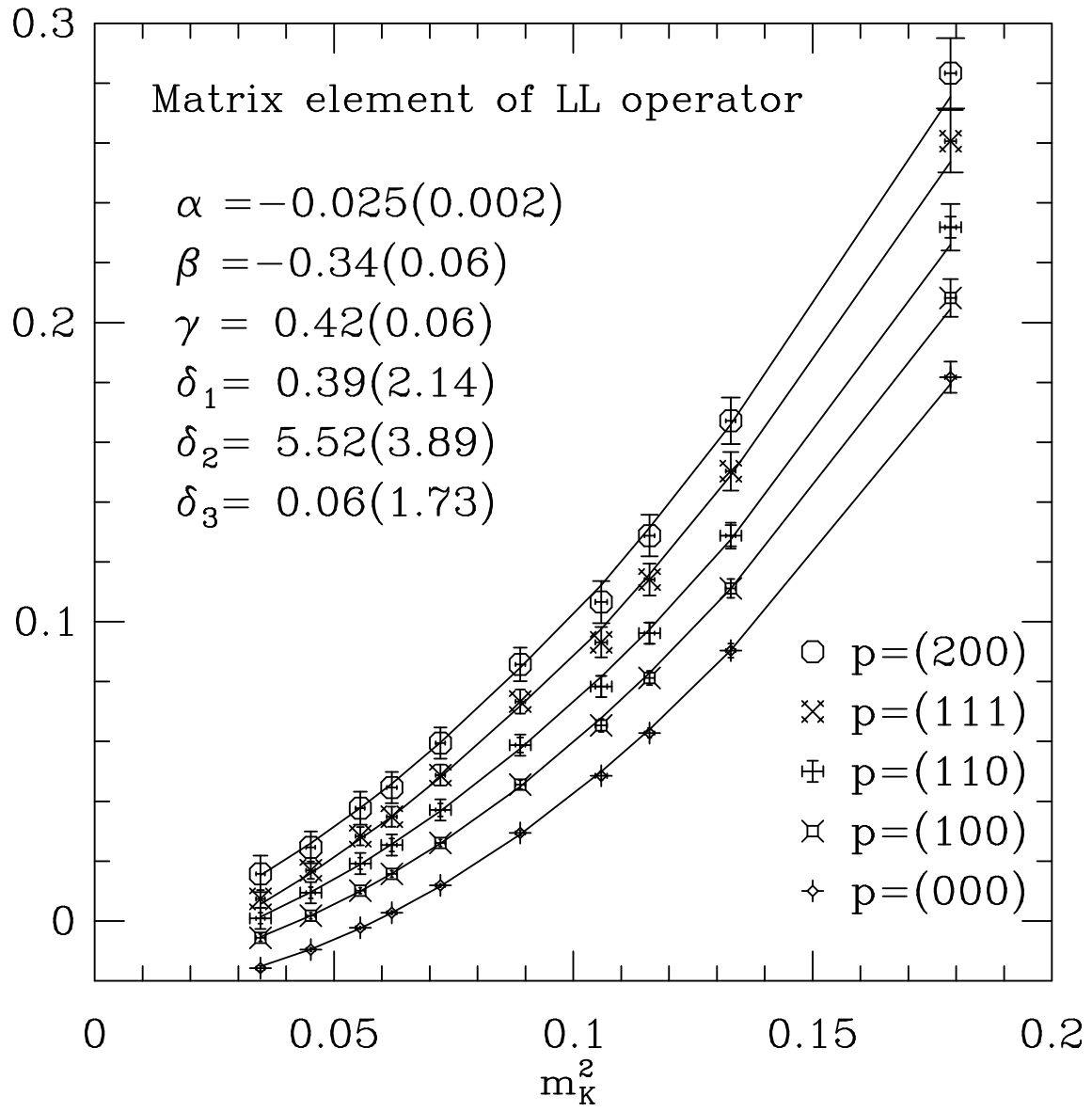

with different momentum transfers $q$:

$$\big(E_1 B_K(q_1) - E_2 B_K(q_2)\big)/(E_1 - E_2) = \gamma + \delta_2 m^2 + \delta_3 m(E_1 + E_2).$$

We correct the term $\delta_3 m(E_1 + E_2)$ using the value of $\delta_3$ extracted from the fit. The results of this analysis for the 10 light mass combinations are given in the third column of Table 5. Interpolating to $m_K$, we get $B_K = 0.68(12)$.

The general form of Eq. 5 follows from Lorentz symmetry as $m^2$ and $p_i{\cdot}p_f$ are the only invariants. The theoretical analysis of what terms are artifacts in heavy-light mesons is not yet complete; indications are that all terms contribute. (The statement that $\alpha, \beta, \delta_1$ are lattice artifacts relies on $\chi PT$ as applied to light-light systems only.) Since data for $CU_i$ show almost no variation with momentum transfer, we therefore extrapolate the unsubtracted value to $\kappa_l$ and get $B_D = 0.73(2)$.

The values of $B_7$ and $B_8$ at the kaon mass are

$$B_7 = 0.76(1), \qquad B_8 = 0.81(1). \tag{7}$$

One loop running gives $B_8(2\ GeV) \approx 0.67$ due to the large anomalous dimension of $O_8$. These estimates are significantly lower than those found in Ref. [6], and the difference is due to a change in lattice size and in the behavior versus $m_q$. Phenomenologically, a value smaller than 1.0 for $B_8$ raises the estimate for $\varepsilon'/\varepsilon$.

Table 5. $B_K$, $B_7$ and $B_8$

| | $B_K$ | | | $B_7$ | $B_8$ |
|---|---|---|---|---|---|
| | $(p=0)$ | $(p=2)$ | subtr. | | |
| $CC$ | 0.87(1) | 0.88(2) | | 0.93(1) | 0.96(1) |
| $CS$ | 0.79(1) | 0.81(2) | | 0.92(1) | 0.95(1) |
| $CU_1$ | 0.77(1) | 0.78(3) | | 0.91(1) | 0.95(1) |
| $CU_2$ | 0.75(1) | 0.76(3) | | 0.91(2) | 0.94(2) |
| $CU_3$ | 0.74(2) | 0.74(4) | | 0.90(2) | 0.94(2) |
| $SS$ | 0.54(1) | 0.61(2) | 0.79(35) | 0.87(1) | 0.91(1) |
| $SU_1$ | 0.44(0) | 0.55(2) | 0.76(27) | 0.84(1) | 0.89(1) |
| $SU_2$ | 0.38(1) | 0.51(3) | 0.73(24) | 0.83(1) | 0.88(1) |
| $SU_3$ | 0.34(1) | 0.47(3) | 0.70(22) | 0.82(1) | 0.87(1) |
| $U_1U_1$ | 0.26(1) | 0.45(3) | 0.72(19) | 0.82(1) | 0.86(1) |
| $U_1U_2$ | 0.14(1) | 0.39(3) | 0.70(16) | 0.80(1) | 0.84(1) |
| $U_1U_3$ | 0.04(1) | 0.34(4) | 0.68(15) | 0.78(1) | 0.83(1) |
| $U_2U_2$ | −0.04(1) | 0.32(5) | 0.69(14) | 0.78(1) | 0.82(1) |
| $U_2U_3$ | −0.21(1) | 0.26(6) | 0.68(12) | 0.76(1) | 0.81(1) |
| $U_3U_3$ | −0.49(2) | 0.20(8) | 0.67(11) | 0.74(1) | 0.79(1) |

## 5. SEMILEPTONIC FORM FACTORS

Results for the form factors for $D \to K e\nu$, $D \to \pi e\nu$, $D \to K^* e\nu$, $D \to \rho e\nu$ decays are given in Tables 6 and 7. Similar results have also been presented by the APE [9], UKQCD [10] and Wuppertal [11] collaborations at this meeting. The details of our method are given in Ref. [8] [4] and the main features of our analysis are:

- The $D$-meson is created at zero spatial momentum and the momentum of the final state meson varies from 0 to $\pi/8a$. This provides a large enough range in the invariant mass of the leptonic subsystem $(-Q^2)$ to test the pole dominance hypotheses. To extract form-factors at $Q^2 = 0$, we make two kinds of fits using $f(Q^2) = f(0)/(1 - Q^2/M^2)$. In the "pole" fits we use the lattice measured value for $M$, while in "best" fits both $f(0)$ and $M$ are free parameters.
- Either LS or SS 2-point correlators can be used to cancel the matrix elements at the source and sink, and the exponential fall off in time as explained in Ref. [8]. Our data show that the meson energies obtained from LS and SS correlators agree in all cases except $p = (1,1,0)$ and $(2,0,0)$ pions. In these cases the $2\sigma$ difference in $E$ (or equivalently the amplitude) translates into a $2\sigma$ difference in $f_+$ and $f_0$. In the vector decay channels the agreement is much better.

Table 6. Semi-leptonic form factors at $Q^2 = 0$

| | $D \to K, K^*$ | | | |
|---|---|---|---|---|
| | $m_s(m_K)$ | | $m_s(m_\phi)$ | |
| | pole | best | pole | best |
| $f_+$ | 0.79(03) | 0.71(05) | 0.79(03) | 0.72(04) |
| $f_0$ | 0.72(02) | 0.73(03) | 0.73(02) | 0.74(03) |
| $f_0/f_+$ | 0.93(01) | 1.02(04) | 0.94(01) | 1.02(03) |
| $V$ | 1.33(07) | 1.33(10) | 1.35(07) | 1.34(09) |
| $A_0$ | 0.86(04) | 0.86(04) | 0.86(04) | 0.87(04) |
| $A_1$ | 0.68(03) | 0.75(05) | 0.70(03) | 0.77(04) |
| $A_2$ | 0.55(15) | 0.64(19) | 0.58(14) | 0.66(18) |
| $A_3$ | 0.86(05) | 0.86(05) | 0.87(04) | 0.87(04) |
| $V/A_1$ | 1.82(08) | 1.75(10) | 1.81(07) | 1.75(09) |
| $A_0/A_1$ | 1.15(07) | 1.15(08) | 1.14(07) | 1.13(07) |
| $A_2/A_1$ | 0.76(17) | 0.85(23) | 0.78(15) | 0.87(21) |
| $A_3/A_1$ | 1.13(08) | 1.15(09) | 1.11(07) | 1.13(08) |

Table 7. Semi-leptonic form factors at $Q^2 = 0$

| | $D \to \pi, \rho$ | | $D_s \to \phi$ |
|---|---|---|---|
| | pole | best | |
| $f_+$ | 0.70(04) | 0.56(08) | |
| $f_0$ | 0.61(03) | 0.62(05) | |
| $f_0/f_+$ | 0.91(02) | 1.07(09) | |
| $V$ | 1.20(11) | 1.18(15) | 1.36(4) |
| $A_0$ | 0.83(06) | 0.82(06) | 0.80(2) |
| $A_1$ | 0.59(03) | 0.67(07) | 0.67(1) |
| $A_2$ | 0.39(21) | 0.44(24) | 0.47(6) |
| $A_3$ | 0.81(06) | 0.81(06) | 0.79(2) |
| $V/A_1$ | 1.89(13) | 1.77(16) | 2.01(4) |
| $A_0/A_1$ | 1.29(12) | 1.23(14) | 1.19(3) |
| $A_2/A_1$ | 0.65(26) | 0.67(31) | 0.68(7) |
| $A_3/A_1$ | 1.23(14) | 1.22(14) | 1.15(3) |

- The $Q^2 = 0$ point lies between momentum transfers $p = (1,1,0)$ and $p = (2,0,0)$. Thus, the estimates of the form-factors at $Q^2 = 0$ are sensitive to data at large momentum transfers.
- $f_+$, $A_1$ and $A_2$ at $Q^2 = 0$ show a significant difference between "pole" and "best" fits. The data prefer values of $M$ which are smaller than that measured from 2-point correlators. Since $m_2 > m_1$, using $m_2$ does not reduce the difference; it increases both estimates by about 5%. We take the "best" fit as our preferred value.
- The $p$ dependence in $Z_A$ and $Z_V$ ($O(pa)$ effects) could account for the difference between "pole" and "best" fits. Therefore, the Lepage-Mackenzie[3] tadpole improved $Z's$ need to be extended to non-zero momentum transfers.
- In order to consistently take into account all $O(ma)$ effects one needs to derive the relationship between matrix elements and form-factors using the lattice symmetries. We have not done that and therefore present all results based on the continuum expressions. A check of this assumption is that the large $t$ behavior of the ratio of the 2-point functions, $\langle V_i V_i \rangle / \langle V_j V_j \rangle$ where $V_i, V_j$ are the spatial components of the local vector current, is $(m^2 + p_i^2)/(m^2 + p_j^2)$. The data verify this at the 2% level for all mass combinations and momenta.
- Our data shows that $A_0(0) \approx A_3(0)$ as predicted by Lorentz invariance.
- In most cases the statistical errors are larger than the difference due to the choice of $m_s$. Our preferred values are with $m_s$ from $m_\phi$.
- We can use the $CU_1 \to U_1 U_1$ measurement to estimate the $D_s \to \phi$ form-factors, assuming that the Zweig suppressed hairpin diagram can be ignored. ($U_1$ mass is very close to $m_s(m_K)$.) The results given in Table 7 show that the $D \to K^*$ and $D_s \to \phi$ form factors are very similar.

To summarize, our estimates of form-factors for charm decays are in good agreement with experimental data. Systematic errors due to using the quenched approximation and due to $O(a)$, $O(pa)$, and $O(m_q a)$ errors remain to be understood and quantified. To extend the analysis to the phenomenologically interesting case of $B$ mesons we propose to combine NRQCD $b$ quarks with existing light propagators.

## 6. THE RARE DECAY $B \to K^* \gamma$

This decay is characterized by three form factors, $T_1$, $T_2$, and $T_3$. Of these only $T_1(0) = T_2(0)$ can be measured in experiments. We use 3 methods to extrapolate our $C$ quark data to $m_b$ using scaling relations from heavy quark effective theory (HQET), and to $Q^2 = 0$ using pole dominance.

Figure 6. A typical example of "pole" versus "best" fits to $f_+$ data.

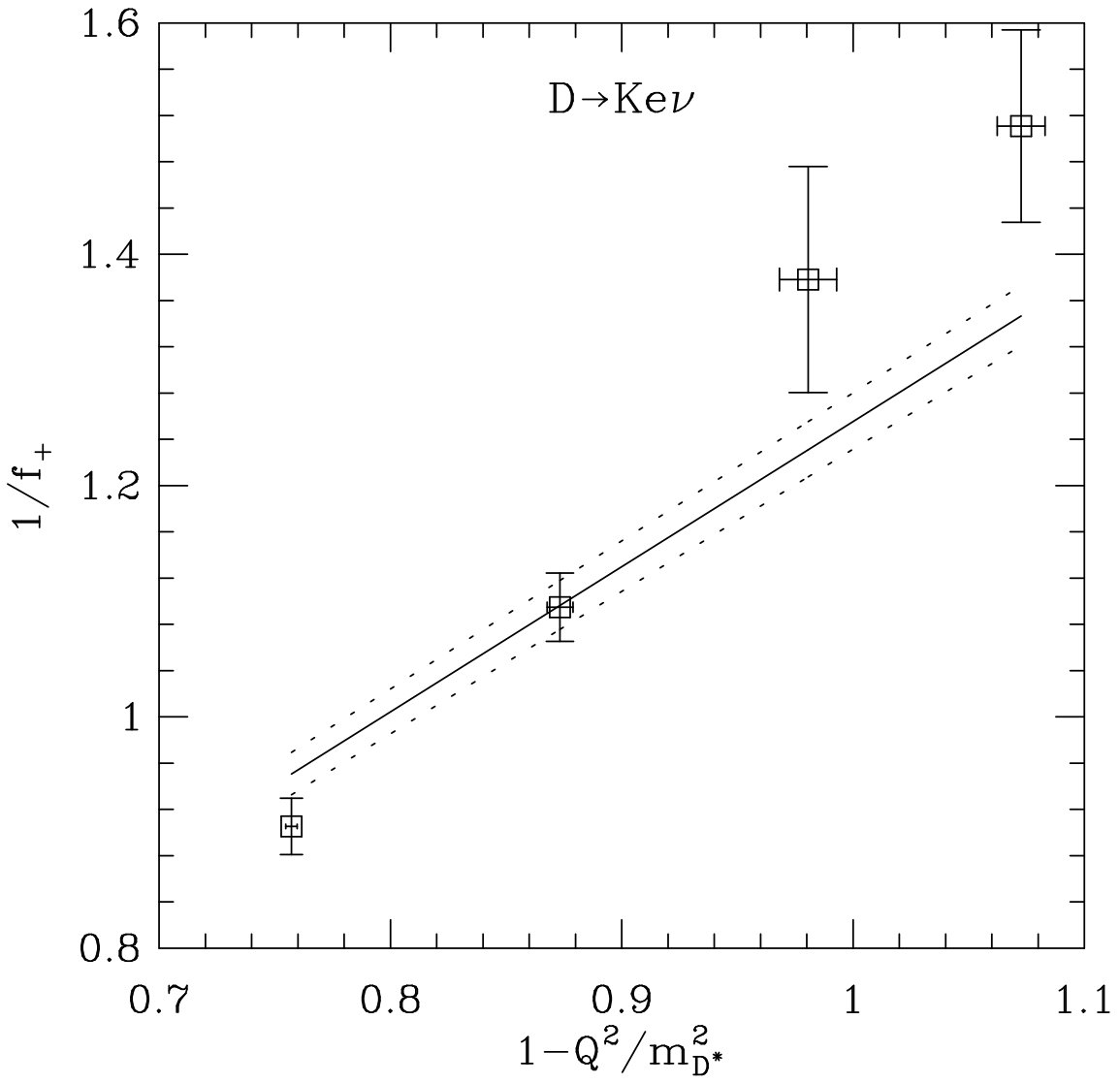

- Method I: $T_2$ can be measured at zero recoil, so we extrapolate the data to the physical B-meson using the HQET relation $T_2(\text{fixed recoil})\ m_B^{1/2} = constant$. We then use pole dominance with the measured value $m_B^* = 5.74$ GeV to estimate $T_2(0) = 0.11(1)$.
- Method II: HQET along with the assumption of pole dominance predicts $T_2(0)m_B^{3/2} = constant$. Therefore we first extrapolate $T_2$ to $Q^2 = 0$ using pole dominance at the charm mass and then using this scaling relation to estimate $T_2(0) = 0.09(1)$ at the $B$ mass.
- Method III: HQET and the pole dominance hypothesis also predicts $T_1(0)m_B^{1/2} = constant$. The procedure of Method II then gives $T_1(0) = 0.25(2)$ at the $B$ mass.

Methods I and II agree, whereas Method III gives a much larger value. The contradiction between Methods II and III arises entirely from the extrapolation to $b$ mass; the lattice values $T_1(0) = 0.39(2)$ and $T_2(0) = 0.41(3)$ are in agreement with the theoretical prediction, based on Lorentz invariance, that $T_1(0) = T_2(0)$ for all values of $m_q$. The problem could be either in the pole dominance hypotheses or in using the lowest order HQET relations. The data in Figure 7 show that $T_1$ prefers a smaller value for $m^2_{D_s^*}$ while $T_2$ a larger one. The difference between "pole" and "best" fits, while small at $c$ mass, could be large at the $b$ mass. The corrections to HQET can be large considering the factor of $2-4$ change in $T$ due to the extrapolation. To summarize, these possibilities need to be brought under control before reliable results for $B \to K^*\gamma$ can be obtained.

Figure 7. Comparison of $T_1$ and $T_2$ data with pole dominance expectation.

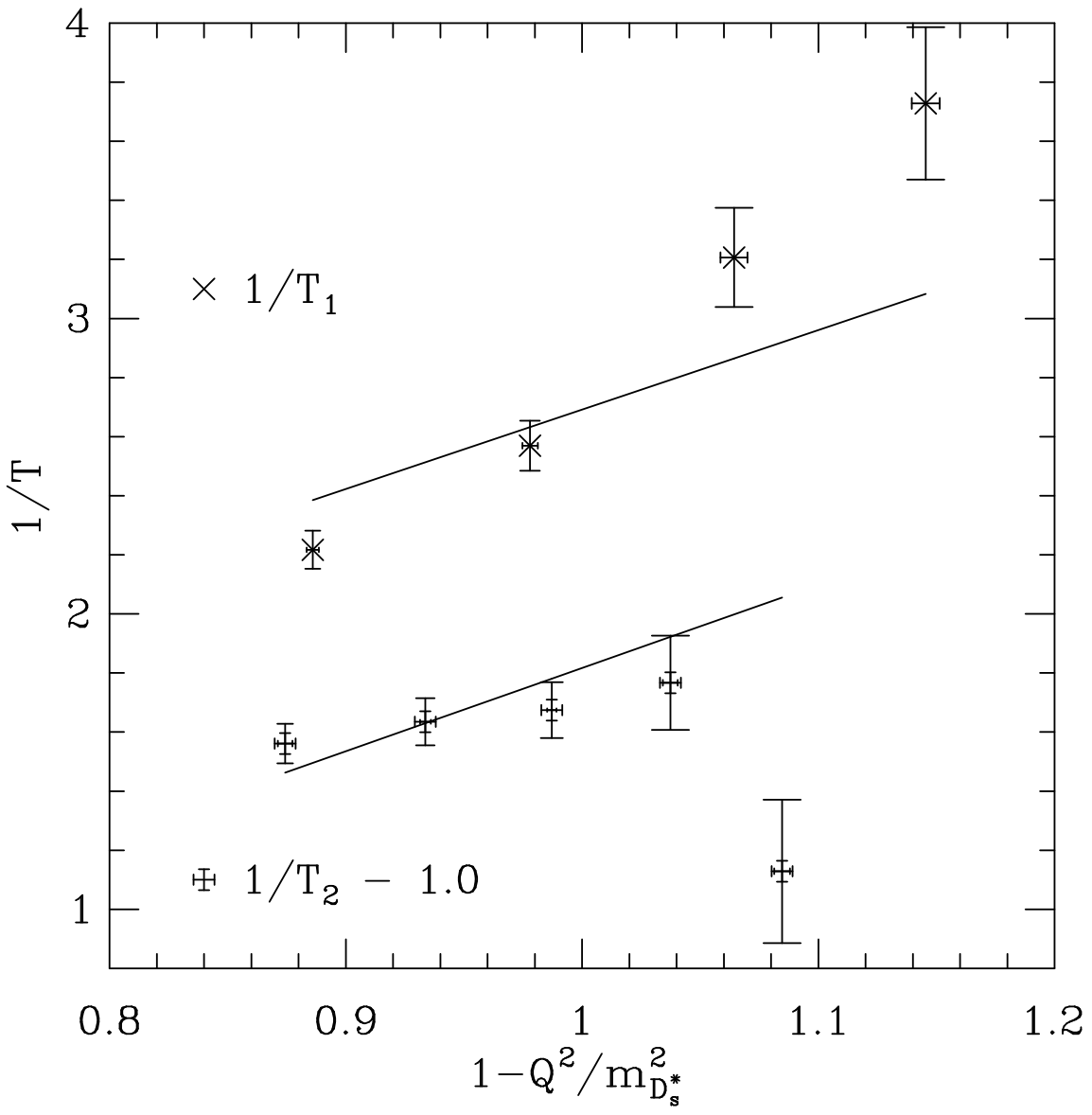

## REFERENCES

1 R. Gupta et al., Phys. Rev. D46 (1992) 3130.
2 R. Gupta, ibid, hep-lat/9412058.
3 P. Lepage, P. Mackenzie, Phys. Rev. D48 (1993) 2250.
4 T. Bhattacharya et al., in preparation.
5 GF11 coll., hep-lat/9405003, hep-lat/9310009
6 R. Gupta et al., Phys. Rev. D47 (1993) 5113.
7 Kilcup et al., Phys. Rev. Lett. 64 (1990) 25.
8 R. Gupta et al., hep-lat/9310007.
9 APE collaboration, ibid, hep-ph/9411011.
10 J. Nieves, ibid, hep-lat/9412013.
11 S. Güsken, ibid, hep-lat/9501013.